# Direct Healthy Life Expectancy Estimates from Life Tables with a Sullivan Extension. Bridging the Gap Between HALE and Eurostat Estimates

**Christos H Skiadas[1] and Charilaos Skiadas[2]**

[1] ManLab, Technical University of Crete, Chania, Crete, Greece
(E-mail: skiadas@cmsim.net )
[2] Department of Mathematics and Computer Science, Hanover College, Indiana, USA
(E-mail: skiadas@hanover.edu )

**Abstract**

The analytic derivation of a more general model of survival-mortality and the estimation of a parameter bx related to the Healthy Life Years Lost (HLYL) is followed with the formulation of a computer program providing results similar to those of the World Health Organization for the Healthy Life Expectancy (HALE) and the corresponding HLYL estimates. This program is an extension of the classical life table including more columns to estimate the cumulative mortality, the average mortality, the person life years lost and finally the HLYL parameter bx. Evenmore, a further extension of the Excel program based on the Sullivan method provides estimates of the Healthy Life Expectancy at every year of the lifespan for five different types of estimates that are the Direct, WHO, Eurostat, Equal and Other. Estimates for several countries are presented. It is also presented a methodology and a program to bridge the gap between the World Health Organization (HALE) and Eurostat (HLE) healthy life expectancy estimates. The latest version of this program (SKI-6 Program) appear in the Demographics2020 website at http://www.smtda.net/demographics2020.html .

## The HLYL Estimation Method

Our methodology Skiadas and Skiadas (2014, 2015, 2018a,b,c, 2019a,b) was based on a geometric approach from the following graph of mortality spaces where both mortality and survival are presented as appropriate areas of this graph.

---

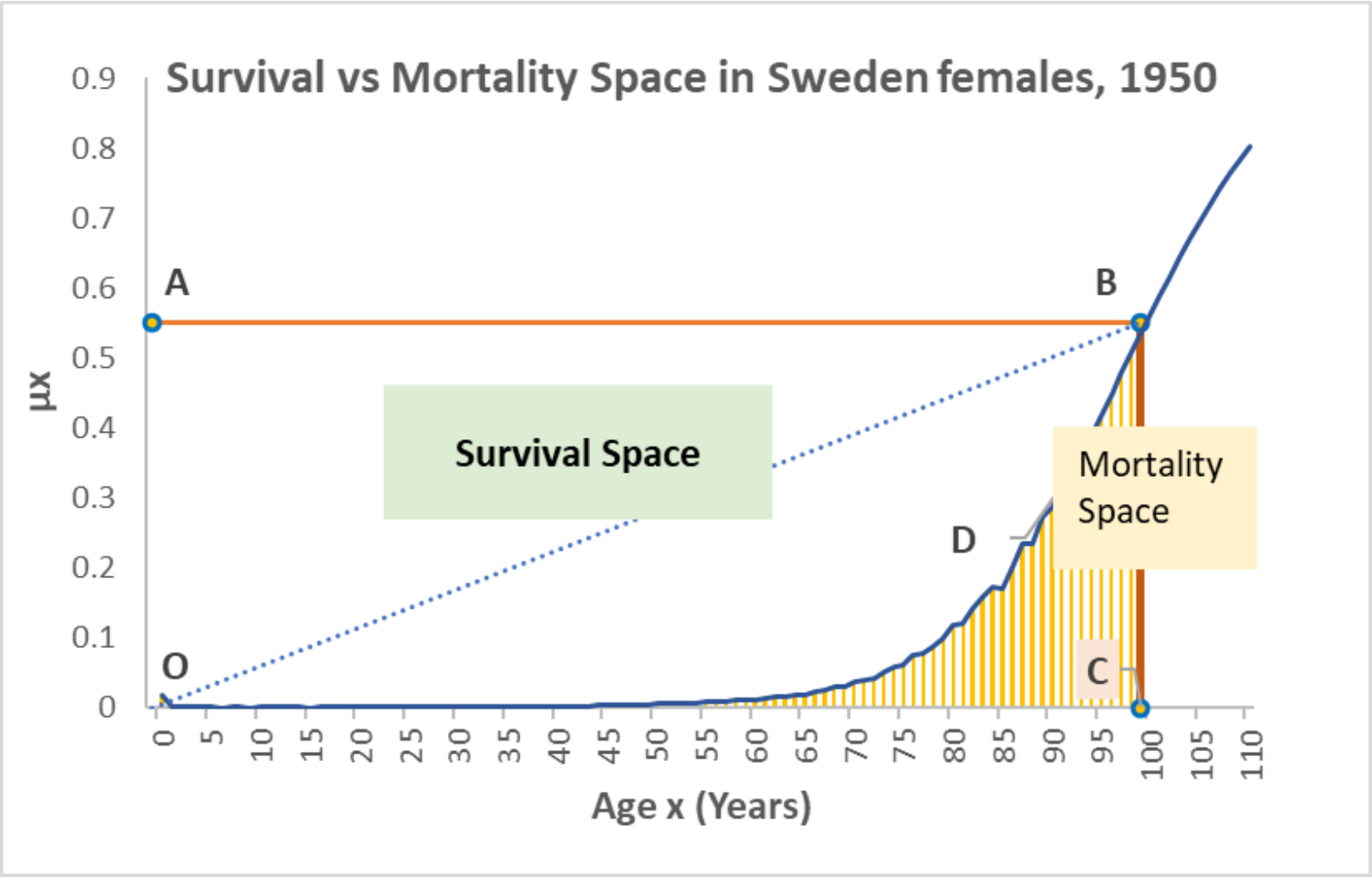

Fig. 1. Survival vs Mortality space graph

The usual form to express mortality $\mu_x$ in a population at age $x$ is by estimating the fraction Death($D_x$)/Population($P_x$) that is $\mu_x$=$D_x$/$P_x$. As in the following we will use the Life Tables provided from the Human Mortality Database we will use the term *mx* of these tables instead of $\mu_x$. The above graph using data from Sweden 1950 females from the HMD is formulated with $\mu_x$ as the blue exponential curve. The main forms of Life Tables start with $\mu_x$ and in the following estimate the survival forms of the population. This methodology leads to the estimation of a probability measure termed as life expectancy at age *x* or life expectancy at birth when considering the total life time. There are several differences between the graph with the survival space above and the survival curves methodology. First of all, the vertical axis in the Survival-Mortality Space (SMS) diagram is the probability $\mu_x$. Instead in the survival diagram the vertical axis represent population (usually it starts from 100.000 in most life tables and gradually slow down until the end). By the SMS diagram we have probability spaces for both survival and mortality. For the age *x* the total space is (ABCOA) in the SMS diagrams that is (OA).(BC)=$x\ \mu_x$. The mortality space is the sum S($\mu_x$) and survival space is ($x\mu_x$ -S($\mu_x$)). Accordingly, the important measure of the Health State is simply the fraction (ABDOA)/(BCODB). Simpler is to prefer the fraction (ABCOA)/(BCODB)=$x\mu_x$ /S($\mu_x$) that can be estimated from $\mu_x$ for every age *x* of the population.

Ruben Roman et al (2007) propose a similar methodology stating: "In the expression of the survival function; H(x) denotes the cumulative hazard function, which is equivalent to the area under the hazard function m(x). The area under the hazard function was defined by taking the corresponding integration limits ranging from x, current age of an individual, to x + $y_x$, age at death or quantity of time lived from birth to death, where X and $Y_x$ are non-negative continuous random variables. The calculated area will give the risk of dying at a given age x up to a particular future time $y_x$". The cumulative hazard they propose is *S* $\mu_x$ where $\mu_x$ is equivalent to the hazard function in our notation.

In modeling the healthy life years lost to disability some important issues should be realized. Mortality expressed by $\mu_x$ is important for modeling disability but more important is the cumulative mortality *S* $\mu_x$ which, as an additive process, is more convenient for the estimation of the healthy life years lived with disability and the deterioration process causing deaths. The estimates for this type of mortality are included in the term $b_x$S $\mu_x$.

Our approach in previous publications (Skiadas and Skiadas (2018a,b,c, 2019)) was to set a time varying fraction $b_x$ for Health/Mortality of the form:

$$b_x = \frac{x\mu_x}{\int_0^x \mu_s ds} \tag{1}$$

This formula is immediately provided from figure 1 by considering the fraction:

$$b_x = \frac{Total\ Space}{Mortality\ Space} = \frac{OABCO}{ODBCO} = \frac{x\mu_x}{\int_0^x \mu_s ds}$$

It should be noted that an alternative approach is given by:

$$b_x = \frac{Survival\ Space}{Mortality\ Space} = \frac{OABCO - ODBCO}{ODBCO} = \frac{x\mu_x}{\int_0^x \mu_s ds} - 1$$

In the latter case the estimated fraction $b_x$ is smaller by one from the previous case. It remains to the applications stage to decide for the most appropriate. So far the Total Space approach is simpler and gave good results.

The main hypothesis is that the population involved in the deterioration process is a fraction of the total population determined by the level of mortality $\mu_x$ at age $x$. Accordingly the mortality process will have two alternatives expressed by the simple equation:

$$x\mu_x = b_x \int_0^x \mu_s ds \approx b_x \sum_0^x \mu_x \tag{2}$$

Where $x\mu_x$ is the incoming part related to the disability of the living population and the second part is the outgoing part that is summed to the mortality for the period from 0 to age $x$. The parameter $b_x$ is a corresponding adding to express the rate of healthy life lost to disability. The applications verify that the maximum values for $b$=$b_{max}$ are compatible to the estimates of the WHO for several countries. Evenmore, our estimates expressing the values for $b_x$ in all the life time are of particularly importance in the studies related to the Health Expenditure estimation.

Some important properties of the last formula are given below:

First we can formulate the Survival Probability $S(t)$

$$S(t) = \exp\left(-\int_0^x \mu_s ds\right) = \exp\left(-\frac{x\mu_x}{b_x}\right) \approx \exp\left(-\sum_0^x \mu_x\right) \tag{3}$$

$$S(t) \approx \exp\left(-\sum_0^x \mu_x\right) = \exp(-\mu_0)\exp(-\mu_1)\exp(-\mu_2)\dots\exp(-\mu_x)$$

Next we can differentiate (2) to obtain

$$(x\mu_x)' = b_x' \int_0^x \mu_s ds + b_x\mu_x \tag{4}$$

For a constant $b$ we have $b_x' = 0$ and

$$x\mu_x' + \mu_x = b\mu_x$$

It follows

$$x\mu'_x = (b-1)\mu_x$$

And rearranging

$$\frac{\mu'_x}{\mu_x} = \frac{b-1}{x}$$

Solving the differential equation

$$\ln(\mu_x) = \ln(c) + (b-1)\ln x$$

Where c is a constant of integration. Finally

$$\mu_x = cx^{b-1}$$

By setting *c=λb* the hazard function or the generating function of the Weibull appear

$$\mu_x = \lambda b x^{b-1} \quad (5)$$

And the cumulative hazard of the Weibull is

$$\Lambda(x) = \lambda x^b = \frac{x\mu_x}{b} = \int_0^x \mu_s ds \quad (6)$$

This is to verify the formula for the survival probability (3) presented earlier.

As we already have presented in previous studies (Skiadas and Skiadas (2018a,b,c, 2019)), $b_x$ can be estimated directly from the life table data. The estimates with the direct method are close to the WHO estimates. The results verify that both methods approach well between each other. Of course the Direct method, based on only the life tables can used in all the time periods as far as life tables exist.

**Program for the Estimates**

We have developed an Excel program for the Direct Estimates of $b_x$ which is provided free of charge. One version can be downloaded from the Demographics 2019 Workshop website at www.asmda.es . The program uses the full life tables from the human mortality database to provide the Healthy Life Year Lost estimator $b_x$ from the general equation form (1):

$$b_x = \frac{x\mu_x}{\int_0^x \mu_s ds}$$

The Cumulative Mortality Mx is given by

$$M_x = \int_0^x \mu_s ds \approx \sum_0^x \left(\frac{dx}{lx}\right) \quad (7)$$

Where dx expresses the death population at age x in the life tables of the HMD and lx is the remaining population at age x in the same life tables. Note that the starting population at age x=0 is set at 100000.

The average mortality Mx/x is estimated by

$$\bar{M}_x = \frac{M_x}{x} \approx \frac{\sum_0^x \left(\frac{dx}{lx}\right)}{x}$$

Then the Person Life Years Lost (PLYL) are provided by

$$PLYL = \frac{dx}{\bar{M}_x} = \frac{xdx}{M_x}$$

The final estimate for $b_x$ is given by

$$b_x = \frac{x\mu_x}{\int_0^x \mu_s ds} \approx \frac{PLYL}{lx} = \frac{xdx}{l_x M_x} = \frac{xdx}{l_x \sum_0^x \left(\frac{dx}{lx}\right)} \quad (8)$$

The methodology is presented in the following figure 2. The full life table from the HMD is followed by 4 more columns for the estimation of $b_x$. In the first, the cumulative mortality is estimated from M=$\sum_0^x \mu_x$. The average mortality $(M/x) = \sum_0^x \mu_x /x$ is provided in the next column whereas the Person Life Years Lost (PLYL)=x$d_x/(\sum_0^x \mu_x)$ are calculated in the following column. Where $d_x$ is provided from the column indicated by dx in the life table. For this very important information an interesting graph is provided. The graph follows a growth process until a high level at 77 years of age and a decline in the remaining lifespan period. It the next column the Healthy Life Year Lost estimator $b_x$ is provided by dividing the PLYL with the lx from the life table. The results are presented in an illustrative graph with the growing trend for $b_x$ to reach a maximum at 9.71 with a decline at higher ages. This high level can be also estimated by fitting the Weibull model (Weibull 1951).

Another option added in this Excel is the estimates of the World Health Organization from 2000-2016 for Life Expectancy at birth and at 60 years of age for all the member countries whereas information for the Healthy Life Expectancy (HALE) at birth and at 60 years of age is provided for the years 2000, 2005, 2010, 2015 and 2016. We have added a small Table to present comparatively the WHO estimates with our estimates with the direct method. The only needed after copy and paste the life table from the HMD to select the name of the country in L1 and the gender (male, female or both sexes) in L2 in the Excel chart. To avoid mistakes we have used list of the WHO countries with their official names.

# Healthy Life Years Lost (HLYL)

Select Country Name: Czechia

Program: C H Skiadas and C Skiadas, 2019, April. Download the Excel files from http://www.asmda.es/demographics2019.html

Select Gender: Female

Manually complete the full Life Table from the Human Mortality Database, HMD (recommended) or from the right side files from this Excel.
Don't forget to Select Country Name and Gender in the appropriate Box

| Year | Age | mx | qx | ax | lx | dx | Lx | Tx | ex | Cumulative mortality Mx=Σ(dx/lx) | Average Cumulative Mortality Mx/x | Years of disability influence at age x | bx HLYL Indicator PLYL/lx | Person life years lost PLYL |
|---|---|---|---|---|---|---|---|---|---|---|---|---|---|---|
| | | | | | | | | | | Mx=Σ(dx/lx) | Mx/x | xdx/Mx | (xdx) / (lxMx) | |
| 2016 | 0 | 0.0025 | 0.00248 | 0.14 | 100000 | 248 | 99788 | 8186842 | 81.87 | 0.002 | 0.010 | 24000 | 0.24 | 1589 |
| 2016 | 1 | 0.0003 | 0.00028 | 0.5 | 99752 | 28 | 99738 | 8087055 | 81.07 | 0.003 | 0.002 | 15214 | 0.17 | 1007 |
| 2016 | 2 | 0.0002 | 0.00015 | 0.5 | 99724 | 15 | 99717 | 7987316 | 80.09 | 0.003 | 0.001 | 12882 | 0.16 | 853 |
| 2016 | 3 | 0.0001 | 0.00013 | 0.5 | 99710 | 13 | 99703 | 7887599 | 79.11 | 0.003 | 0.001 | 14960 | 0.13 | 991 |
| 2016 | 4 | 9E-05 | 0.00009 | 0.5 | 99697 | 9 | 99692 | 7787896 | 78.12 | 0.003 | 0.001 | 12932 | 0.12 | 856 |
| 2016 | 5 | 5E-05 | 0.00005 | 0.5 | 99687 | 5 | 99685 | 7688204 | 77.12 | 0.003 | 0.001 | 8643 | 0.14 | 572 |
| 2016 | 6 | 5E-05 | 0.00005 | 0.5 | 99682 | 5 | 99679 | 7588520 | 76.13 | 0.003 | 0.000 | 10055 | 0.14 | 666 |
| 2016 | 7 | 0.0001 | 0.0001 | 0.5 | 99677 | 10 | 99672 | 7488841 | 75.13 | 0.003 | 0.000 | 22506 | 0.15 | 1490 |
| 2016 | 8 | 7E-05 | 0.00007 | 0.5 | 99667 | 7 | 99663 | 7389169 | 74.14 | 0.003 | 0.000 | 17486 | 0.15 | 1158 |
| 2016 | 9 | 5E-05 | 0.00005 | 0.5 | 99660 | 5 | 99657 | 7289505 | 73.14 | 0.003 | 0.000 | 13757 | 0.18 | 911 |
| 2016 | 10 | 4E-05 | 0.00004 | 0.5 | 99655 | 4 | 99653 | 7189848 | 72.15 | 0.003 | 0.000 | 12024 | 0.20 | 796 |
| 2016 | 11 | 8E-05 | 0.00008 | 0.5 | 99651 | 8 | 99647 | 7090195 | 71.15 | 0.004 | 0.000 | 25747 | 0.27 | 1705 |
| 2016 | 12 | 9E-05 | 0.00009 | 0.5 | 99643 | 9 | 99638 | 6990549 | 70.16 | 0.004 | 0.000 | 30708 | 0.38 | 2033 |
| 2016 | 13 | 0.0002 | 0.00015 | 0.5 | 99634 | 15 | 99627 | 6890910 | 69.16 | 0.004 | 0.000 | 53092 | 0.43 | 3516 |
| 2016 | 14 | 0.0002 | 0.00018 | 0.5 | 99619 | 18 | 99610 | 6791284 | 68.17 | 0.004 | 0.000 | 65335 | 0.51 | 4326 |
| 2016 | 15 | 0.0001 | 0.00011 | 0.5 | 99601 | 11 | 99596 | 6691674 | 67.18 | 0.004 | 0.000 | 41532 | 0.60 | 2750 |
| 2016 | 16 | 0.0002 | 0.00016 | 0.5 | 99590 | 16 | 99582 | 6592078 | 66.19 | 0.004 | 0.000 | 61886 | 0.69 | 4098 |
| 2016 | 17 | 0.0002 | 0.0002 | 0.5 | 99574 | 20 | 99564 | 6492496 | 65.2 | 0.004 | 0.000 | 78357 | 0.71 | 5188 |
| 2016 | 18 | 0.0003 | 0.00025 | 0.5 | 99554 | 24 | 99542 | 6392932 | 64.22 | 0.005 | 0.000 | 94311 | 0.80 | 6245 |
| 2016 | 19 | 0.0002 | 0.0002 | 0.5 | 99529 | 20 | 99520 | 6293390 | 63.23 | 0.005 | 0.000 | 79450 | 0.81 | 5261 |
| 2016 | 20 | 0.0002 | 0.00021 | 0.5 | 99510 | 21 | 99499 | 6193871 | 62.24 | 0.005 | 0.000 | 84085 | 0.81 | 5568 |

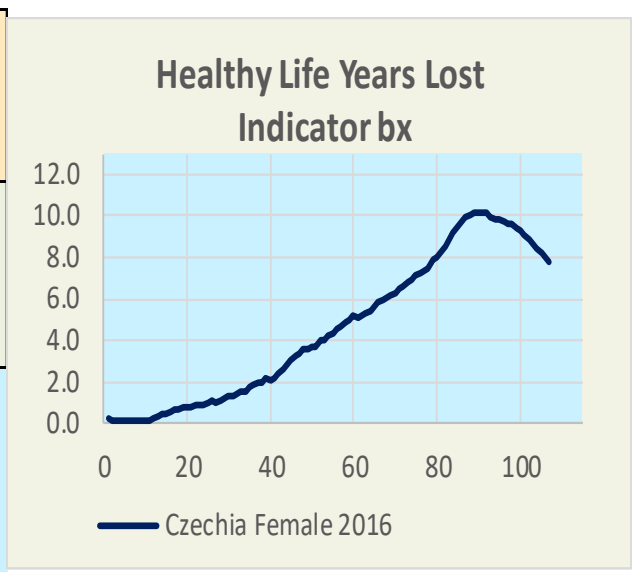

| Parameters | WHO Estimates | Direct HLYL Estimates |
|---|---|---|
| LE | 82.06 | 81.87 |
| HALE | 71.56 | 71.66 |
| HLYL | 10.50 | 10.21 |
| LE at 60 | 24.15 | 24.03 |
| HALE at 60 | 18.18 | 18.91 |
| HLYL at 60 | 5.97 | 5.12 |

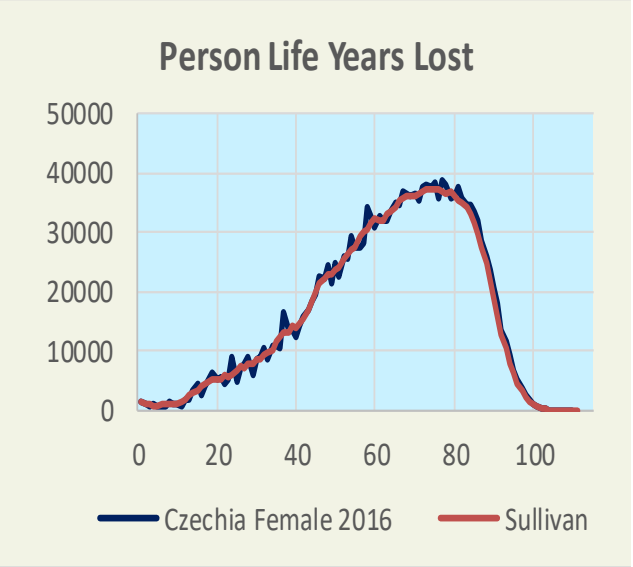

Fig. 2. The extended Life Table for the HLYL estimates. Download full program from http://www.smtda.net/demographics2020.html

| | | | |
|---|---|---|---|
| 26.4282 | 64 | Eurostat HLE | |
| 15.1021 | Select | Parameter | |
| **Sullivan Method** | **Eurostat** | **0.0662** | |

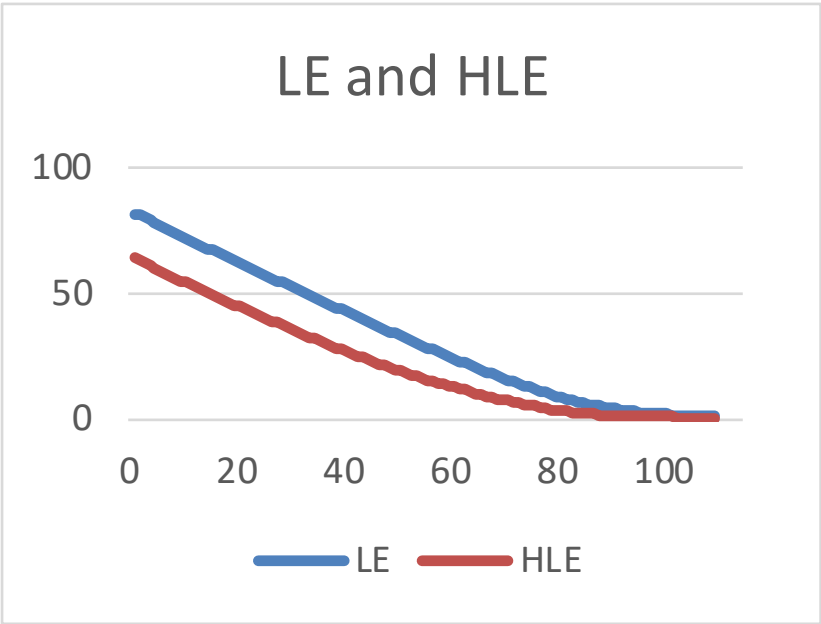

| Proportion with disability | Person Years Lived wthout disability | Total Years Lived without disability | Disability free life expectancy | Healthy Life Years Lost | Index | Health Parameter | Person Years Lost to disability | Person Years Lived wthout disability | Total Years Lived without disability | Disability free life expectancy |
|---|---|---|---|---|---|---|---|---|---|---|
| 0.02 | 98202 | 6400002 | 64.00 | 17.87 | Direct | 0.0378 | 1585.81 | 98199 | 6372712 | 63.73 |
| 0.01 | 98590 | 6301800 | 63.17467 | 17.90 | Equal | 0.0609 | 1148.45 | 98731 | 6274513 | 62.90 |
| 0.01 | 98659 | 6203210 | 62.20379 | 17.89 | Eurostat | 0.0662 | 1058.35 | 98864 | 6175783 | 61.93 |
| 0.01 | 98847 | 6104552 | 61.22306 | 17.89 | Other | 0.055 | 855.78 | 98712 | 6076919 | 60.95 |
| 0.01 | 98904 | 6005704 | 60.23957 | 17.88 | WHO | 0.0389 | 787.51 | 98836 | 5978206 | 59.96 |
| 0.01 | 98770 | 5906800 | 59.25346 | 17.87 | | | 915.01 | 99113 | 5879371 | 58.98 |
| 0.01 | 98730 | 5808030 | 58.26558 | 17.86 | | | 948.50 | 99013 | 5780258 | 57.99 |
| 0.01 | 98713 | 5709299 | 57.278 | 17.85 | | | 959.43 | 98182 | 5681245 | 57.00 |
| 0.01 | 98659 | 5610587 | 56.29333 | 17.85 | | | 1004.16 | 98505 | 5583063 | 56.02 |
| 0.01 | 98445 | 5511928 | 55.30733 | 17.83 | | | 1211.96 | 98746 | 5484558 | 55.03 |
| 0.01 | 98332 | 5413483 | 54.32224 | 17.83 | | | 1320.64 | 98857 | 5385812 | 54.04 |
| 0.02 | 97855 | 5315151 | 53.33765 | 17.81 | | | 1792.24 | 97942 | 5286955 | 53.05 |
| 0.02 | 97163 | 5217296 | 52.35988 | 17.80 | | | 2475.34 | 97605 | 5189013 | 52.08 |
| 0.03 | 96761 | 5120133 | 51.38942 | 17.77 | | | 2866.04 | 96111 | 5091408 | 51.10 |
| 0.03 | 96266 | 5023372 | 50.42584 | 17.74 | | | 3344.46 | 95284 | 4995297 | 50.14 |
| 0.04 | 95621 | 4927107 | 49.46845 | 17.71 | | | 3975.43 | 96846 | 4900013 | 49.20 |
| 0.05 | 95061 | 4831486 | 48.51377 | 17.68 | | | 4521.46 | 95484 | 4803167 | 48.23 |
| 0.05 | 94856 | 4736426 | 47.56689 | 17.63 | | | 4708.41 | 94376 | 4707683 | 47.28 |
| 0.05 | 94270 | 4641570 | 46.62364 | 17.60 | | | 5271.59 | 93297 | 4613307 | 46.34 |
| 0.05 | 94206 | 4547300 | 45.68819 | 17.54 | | | 5314.41 | 94259 | 4520010 | 45.41 |
| 0.05 | 94188 | 4453094 | 44.75022 | 17.49 | | | 5311.34 | 93931 | 4425751 | 44.48 |

Fig. 3. The extended Life Table for the HLYL estimates with the Sullivan method

## A further HLE estimate based on the Sullivan method

The classical Sullivan method is a standard devise to estimate the healthy life expectancy (see Sullivan, 1971 and Jagger, Van Oyen and Robine 1999). The simplicity of this method and the possibility to use it as a continuation or extension of the life table gave as the opportunity to add a Sullivan extension to the above extended life table. By this extension we have estimated the Healthy Life Years Lost (HLYL) for all the life span along with the Healthy Life Expectancy (HLE). The main part of the estimate is based on the proportion lived with disability. This is created form the bx indicator from the previous columns multiplied with a discount Health Parameter. This is estimated directly from the program for our “direct” estimates and for WHO as well. For the Eurostat estimates it is necessary to add the Healthy Life Expectancy at Birth in the appropriate box in the top of Figure 3. Another opportunity is by selecting the “Equal” option which automatically provides an estimate higher than HALE and closer to Eurostat. Everyone of these options should be selected manually in the “Select” and “Parameter” places. The “Other” selection is set by the program user if needed. After selecting the appropriate Health Parameter, the next estimates follow automatically the Sullivan method, first for estimating the person years lived without disability, then the total years lived without disability and finally the Disability Free Life Expectancy. It follows the life years lived with disability column. An alternative method based on the Sullivan system is presented in the columns to the right hand side of Figure 3. The estimates are based on the Person Years Lost to Disability and the direct estimates of the person years lived without disability, the total years lived without disability and the Disability Free Life Expectancy with similar results with the previous approach. The estimated Healthy Life Years from two methods are presented in the appropriate graph of Figure 4. The Health Parameters are estimated as follows:

Direct Parameter= sumproduct(Lx.bx)/(Tx-(LE-HLYL).lx)

WHO Parameter= sumproduct(Lx.bx)/(Tx-(HALE).lx)

Eurostat Parameter=sumproduct(Lx.bx)/(Tx-HLE.lx)

The inverse of the Equal Parameter corresponds to the Healthy Life Years Lost, that is:

HLYL=1/Equal

For the application for Czechia males and females in 2016 the related estimates are included in the next Table:

| TABLE I HLE and HLYL for males and females in Czechia in 2016 | | | | |
|---|---|---|---|---|
| Method | HLE-males | HLYL-males | HLE-females | HLYL-females |
| Direct | 67.7 | 8.4 | 71.7 | 10.2 |
| WHO | 67.1 | 9.2 | 71.6 | 10.5 |
| Eurostat | 62.7 | 13.4 | 64.0 | 17.9 |
| Equal | 60.9 | 15.2 | 65.4 | 16.4 |
| Related estimates from the SKI-6 Program | | | | |
| SKI-6 Severe+Light | 60.8 | 15.3 | 61.5 | 20.4 |
| SKI-6 Severe | 68.2 | 7.9 | 71.3 | 10.6 |
| SKI-6 Moderate | 68.7 | 7.4 | 72.1 | 9.8 |

Clearly the estimates for the HLE and the HLYL for males and females are similar for the Direct and the WHO methods. As also is found from the estimates based on the SKI-6 program (see Figure 10) presented in previous publications both estimates, Direct and WHO, are close to the Severe disability cases. Instead, the estimates provided by the Eurostat (Eurostat, 2019) and Equal are close to Severe and Moderate disability cases. To this end, the Direct and WHO estimates refer to an important part of the life span with the life years lost to disability governing the development of the everyday life.

Our Direct method, further from estimating the Healthy Life Expectancy, has the advantage of providing the Healthy Life Years Lost at every year of age via the bx parameter as is illustrated in Figure 4. Males and females show similar behavior for the age period 0-70 years of age with the exception of the years from 17-30 where an excess of life years lost to disability appear with the form of a higher bx for males than for females. After 70 years of age the bx for females become higher with a maximum level at 88 years of age (b=10.21) with a decline for the higher age years. For males the maximum is at 92 years (b=8.37) with a decline at higher ages.

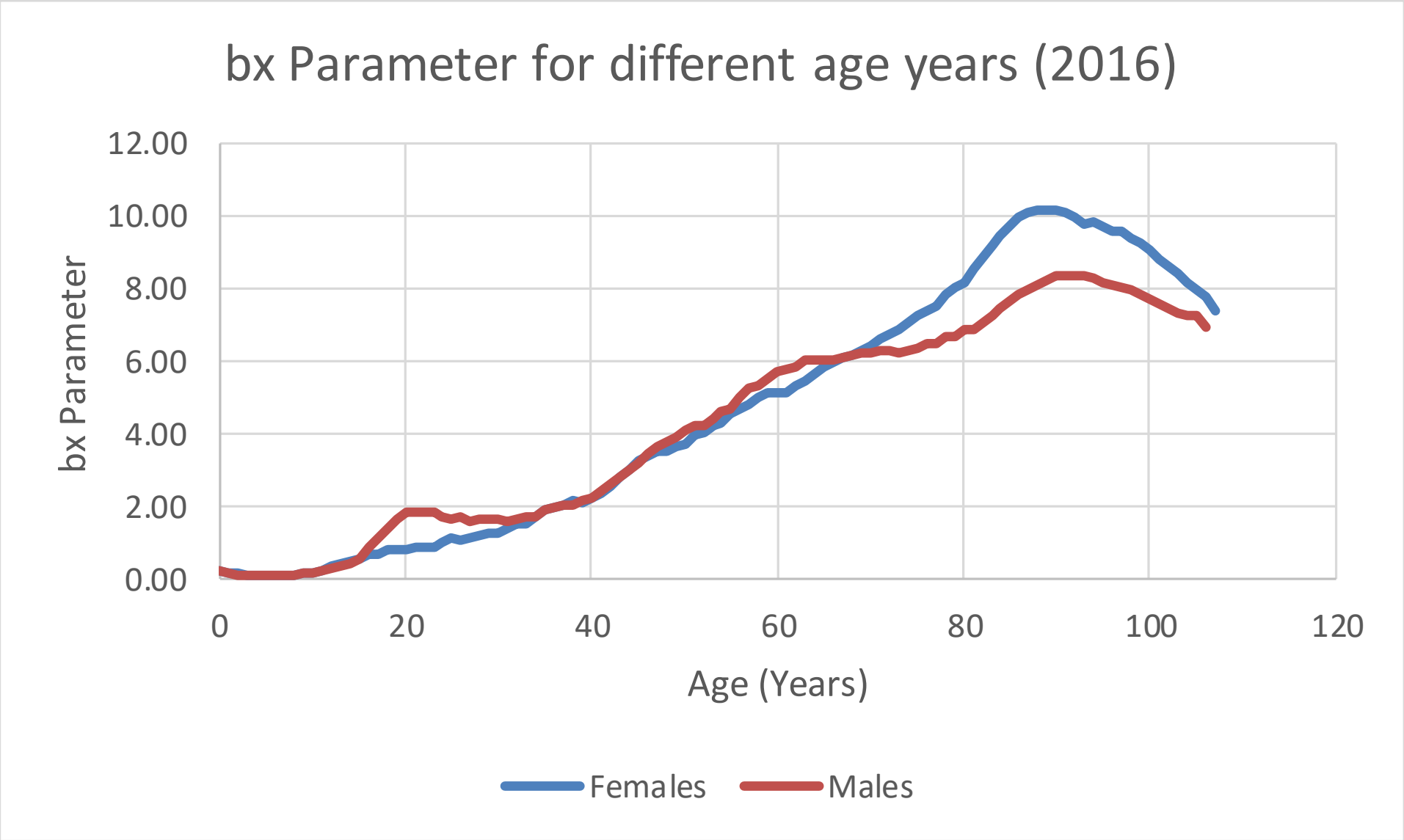

Fig. 4. Healthy Life Years Lost Parameter (bx) for males and females in Czechia in 2016

Note that the proportion with disability for males and females in Czechia in 2016 as constructed from Figure 3 data in the Excel program is illustrated in Figure 5. Similarities appear with the bx estimates presented in Figure 4 with higher proportion with disability for males for the years 50-70 further to the higher values from 17-30 years. The next part for the +70 years includes higher values for females thus explaining the higher Healthy Life Years Lost for females than for males.

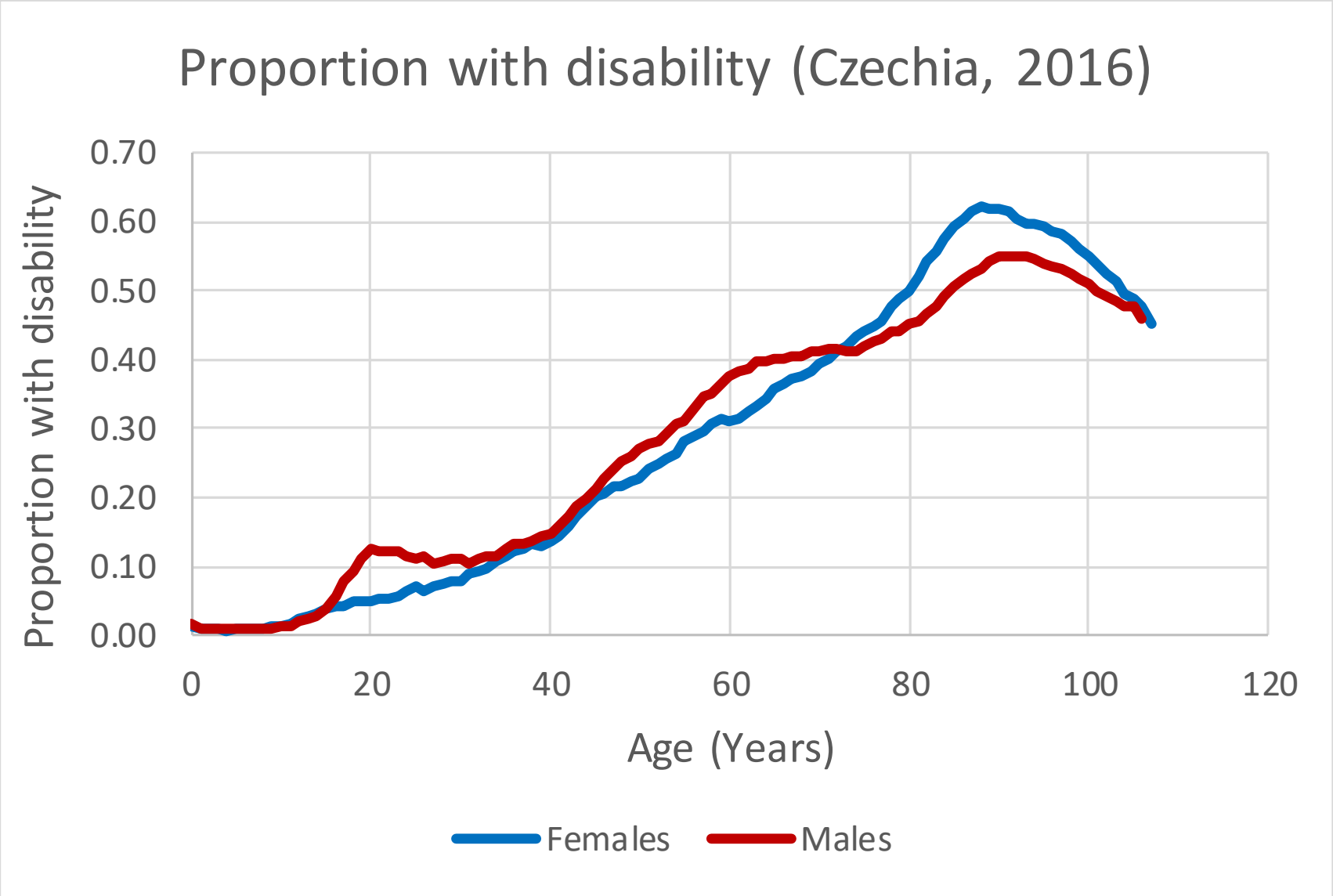

Fig. 5. Proportion with disability for males and females in Czechia in 2016.

**Bridging the Gap between WHO and Eurostat estimates for Healthy Life Years Lost**

Note that the estimates of Healthy Life Years Lost for males and females provided by WHO (2019) and Eurostat (2019) differ considerably in several cases as is illustrated in Figure 6 and Figure 7 due to different methodologies applied. However, in few cases provide similar results namely for Bulgaria, Malta, Norway and Sweden for males and Bulgaria, Malta and Sweden for females (see Figures 6 and 7).

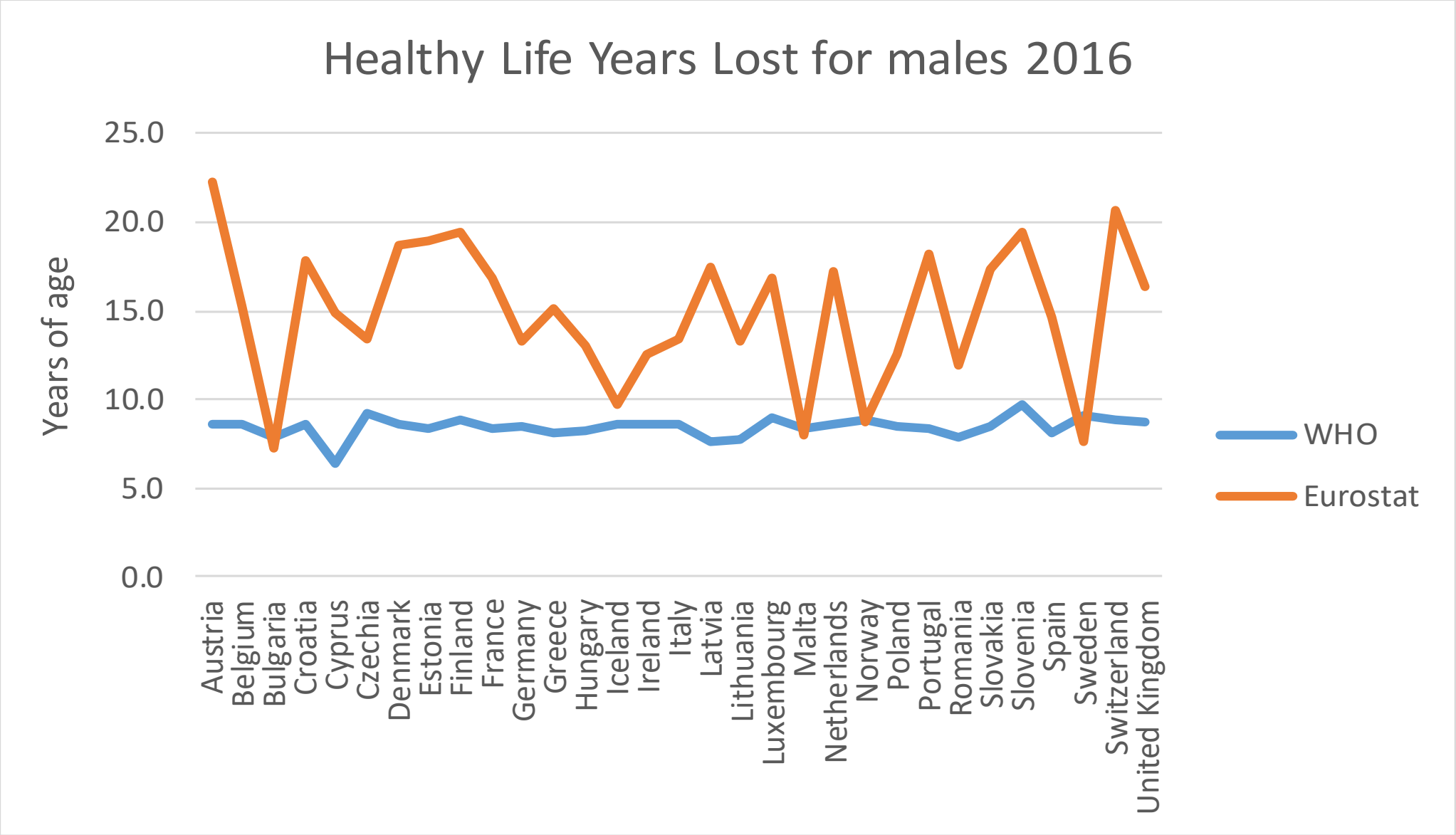

Fig. 6. Healthy Life Years Lost for males in 2016 estimates from WHO and Eurostat.

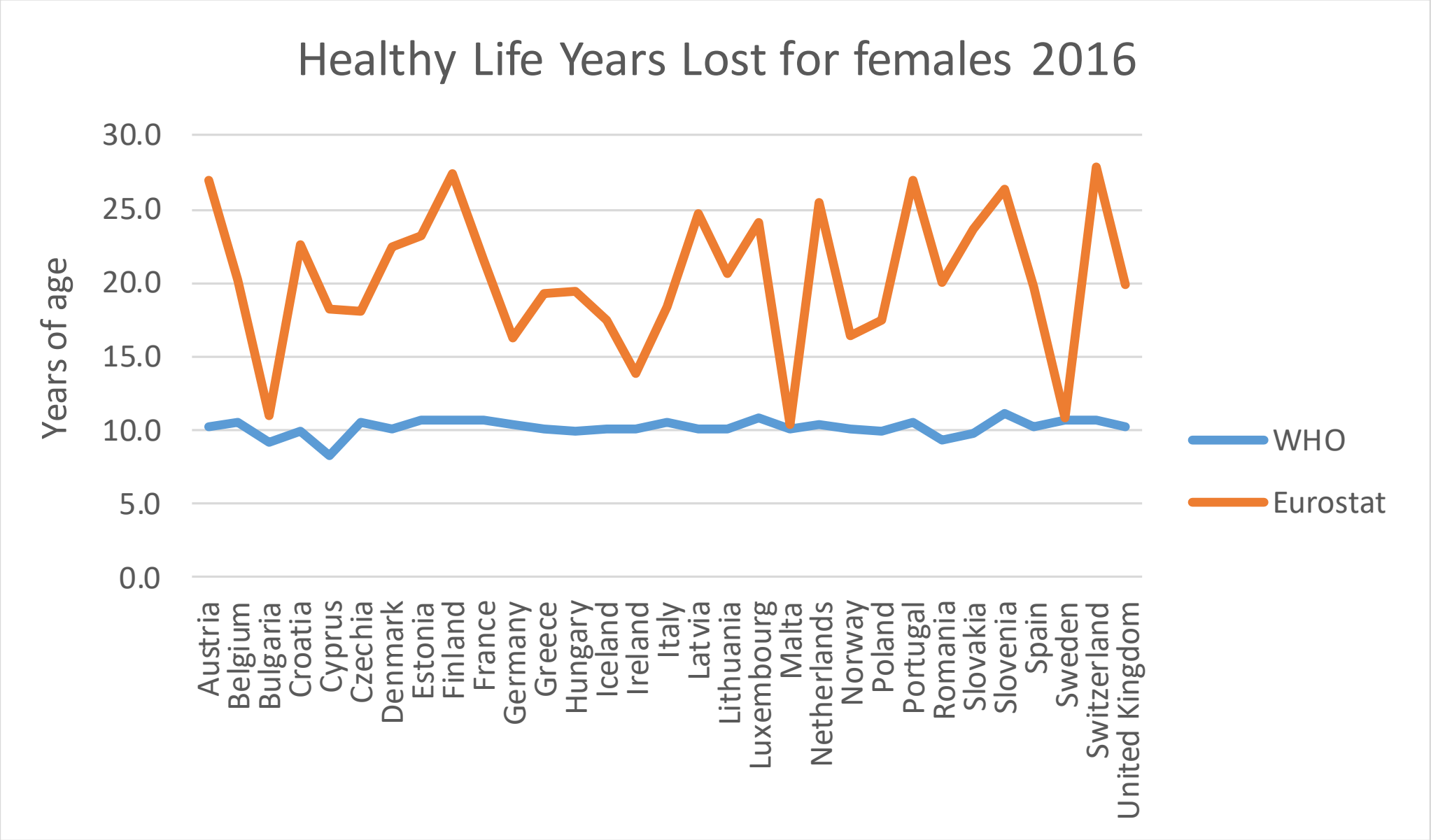

Fig. 7. Healthy Life Years Lost for females in 2016 estimates from WHO and Eurostat.

To bridge the gap between WHO and Eurostat estimates we use the SKI-6 program presented in previous publications providing the figures included in Table II. Healthy Life Years Lost due to severe or moderate disability causes are estimated separately using the SKI-6 Excel Program (Skiadas-Skiadas, 2018, Skiadas_Arezzo, 2018). The estimates for WHO are close to the Severe disability cases. Instead, the estimates provided by the Eurostat are close to the sum of Severe and Moderate disability cases as presented in the same table and illustrated in Figures 8 and 9 for the year 2016 for European countries where the life tables are provided from the Human Mortality Database for the specific year. The results support the argument of two interrelated estimates useful for various health related applications. However, the WHO estimates providing the HLYL from severe disability causes are most valuable when the health expenditure allocation is calculated. Severe disabilities need extensive health resources and intense treatment.

TABLE II. Healthy Life Years Lost to Disability for Males and Females in 2016

| | Females | | | | | Males | | | | |
|---|---|---|---|---|---|---|---|---|---|---|
| Country | WHO | Eurostat | Severe Disability | Moderate Disability | Total Disability | WHO | Eurostat | Severe Disability | Moderate Disability | Total Disability |
| Austria | 10.3 | 27.0 | 12.2 | 10.2 | 22.3 | 8.6 | 22.3 | 9.1 | 9.9 | 19.0 |
| Belgium | 10.5 | 20.2 | 11.3 | 10.8 | 22.1 | 8.6 | 15.3 | 9.1 | 9.8 | 18.9 |
| Croatia | 9.9 | 22.6 | 10.3 | 9.7 | 20.0 | 8.6 | 17.9 | 9.0 | 6.7 | 15.7 |
| Czech Rep. | 10.5 | 18.1 | 10.6 | 9.8 | 20.4 | 9.2 | 13.4 | 7.8 | 7.5 | 15.2 |
| Denmark | 10.1 | 22.5 | 9.1 | 10.7 | 19.7 | 8.6 | 18.7 | 9.1 | 10.2 | 19.3 |
| Estonia | 10.7 | 23.2 | 10.7 | 10.3 | 21.0 | 8.4 | 18.9 | 7.7 | 5.9 | 13.6 |
| France | 10.7 | 21.6 | 11.6 | 12.0 | 23.6 | 8.3 | 16.9 | 9.1 | 12.4 | 21.5 |
| Germany | 10.3 | 16.2 | 11.0 | 11.3 | 22.3 | 8.4 | 13.3 | 9.3 | 10.0 | 19.3 |
| Hungary | 9.9 | 19.5 | 9.1 | 10.1 | 19.2 | 8.2 | 13.1 | 7.1 | 4.5 | 11.7 |
| Latvia | 10.1 | 24.7 | 10.5 | 9.7 | 20.1 | 7.7 | 17.5 | 7.8 | 3.9 | 11.7 |
| Lithuania | 10.2 | 20.7 | 10.6 | 10.6 | 21.1 | 7.7 | 13.3 | 6.8 | 5.4 | 12.3 |
| Netherlands | 10.4 | 25.4 | 11.0 | 11.5 | 22.5 | 8.7 | 17.2 | 9.5 | 9.8 | 19.3 |
| Poland | 9.9 | 17.4 | 9.9 | 10.7 | 20.6 | 8.5 | 12.6 | 6.9 | 6.7 | 13.7 |
| Slovak Rep. | 9.8 | 23.7 | 9.7 | 9.8 | 19.5 | 8.5 | 17.4 | 8.0 | 5.4 | 13.5 |
| Slovenia | 11.1 | 26.4 | 11.2 | 10.8 | 22.0 | 9.7 | 19.5 | 9.1 | 9.0 | 18.1 |
| Spain | 10.3 | 19.8 | 10.5 | 11.5 | 22.0 | 8.1 | 14.6 | 8.9 | 10.5 | 19.3 |
| Sweden | 10.8 | 10.8 | 10.8 | 10.6 | 21.5 | 9.0 | 7.6 | 9.3 | 10.6 | 19.9 |
| Switzerland | 10.7 | 27.9 | 11.2 | 11.4 | 22.6 | 8.8 | 20.7 | 10.3 | 10.3 | 20.6 |
| United Kingdom | 10.3 | 19.9 | 8.9 | 12.0 | 20.9 | 8.7 | 16.4 | 9.1 | 9.6 | 18.7 |

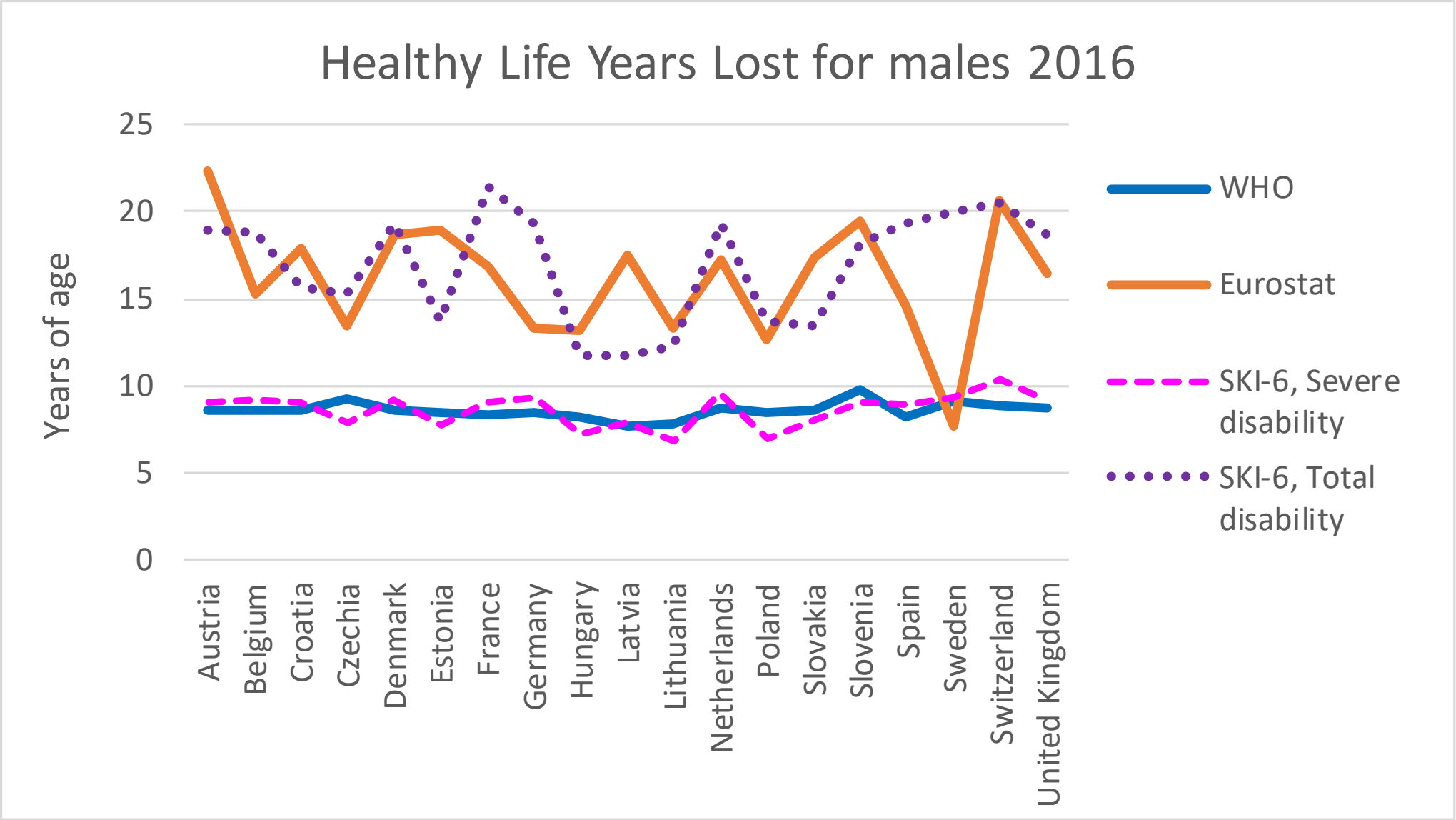

Fig. 8. Healthy Life Years Lost for males in 2016 (WHO and Eurostat and SKI-6 Program).

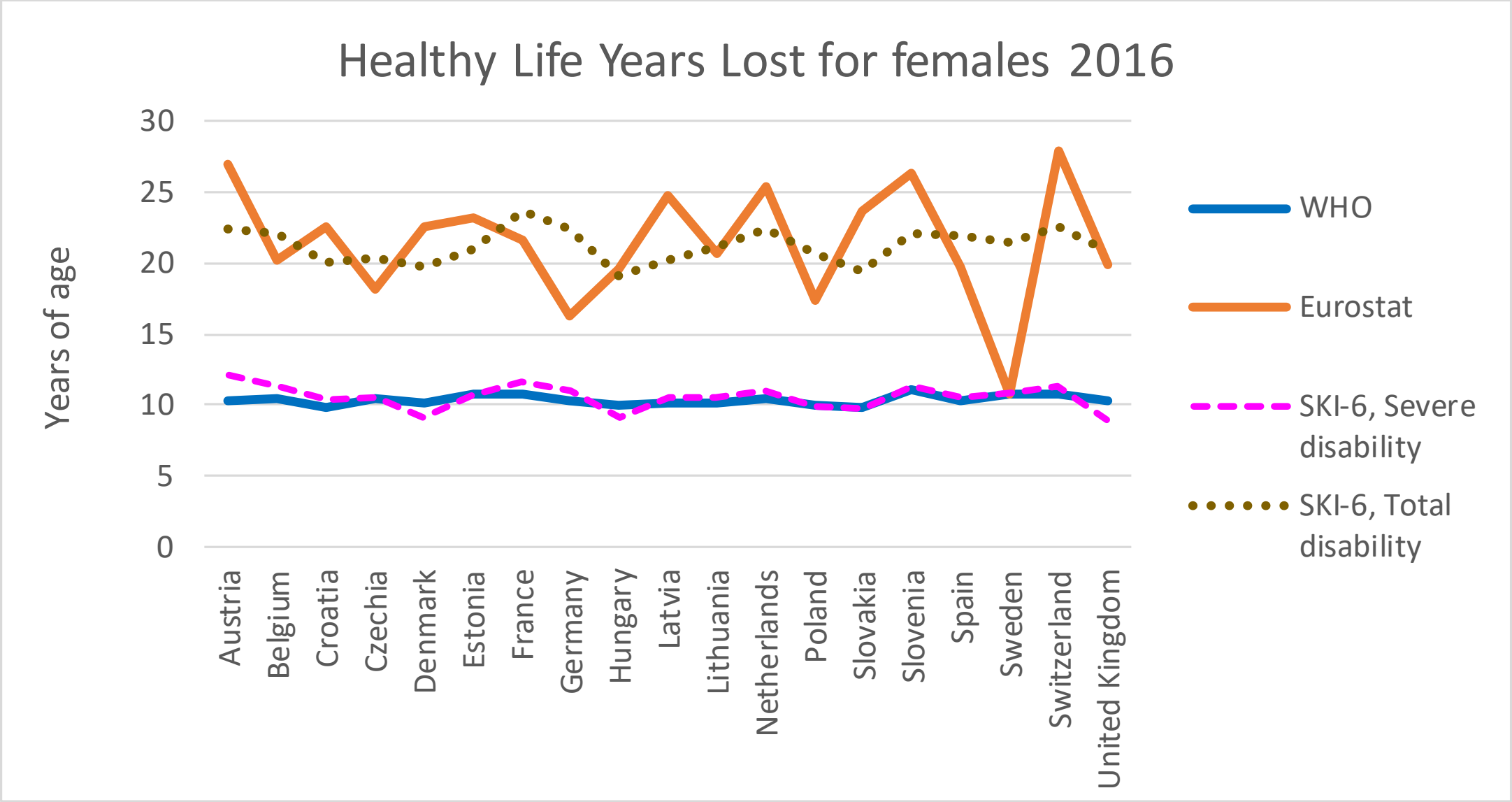

Fig. 9. Healthy Life Years Lost for females in 2016 (WHO and Eurostat and SKI-6 Program).

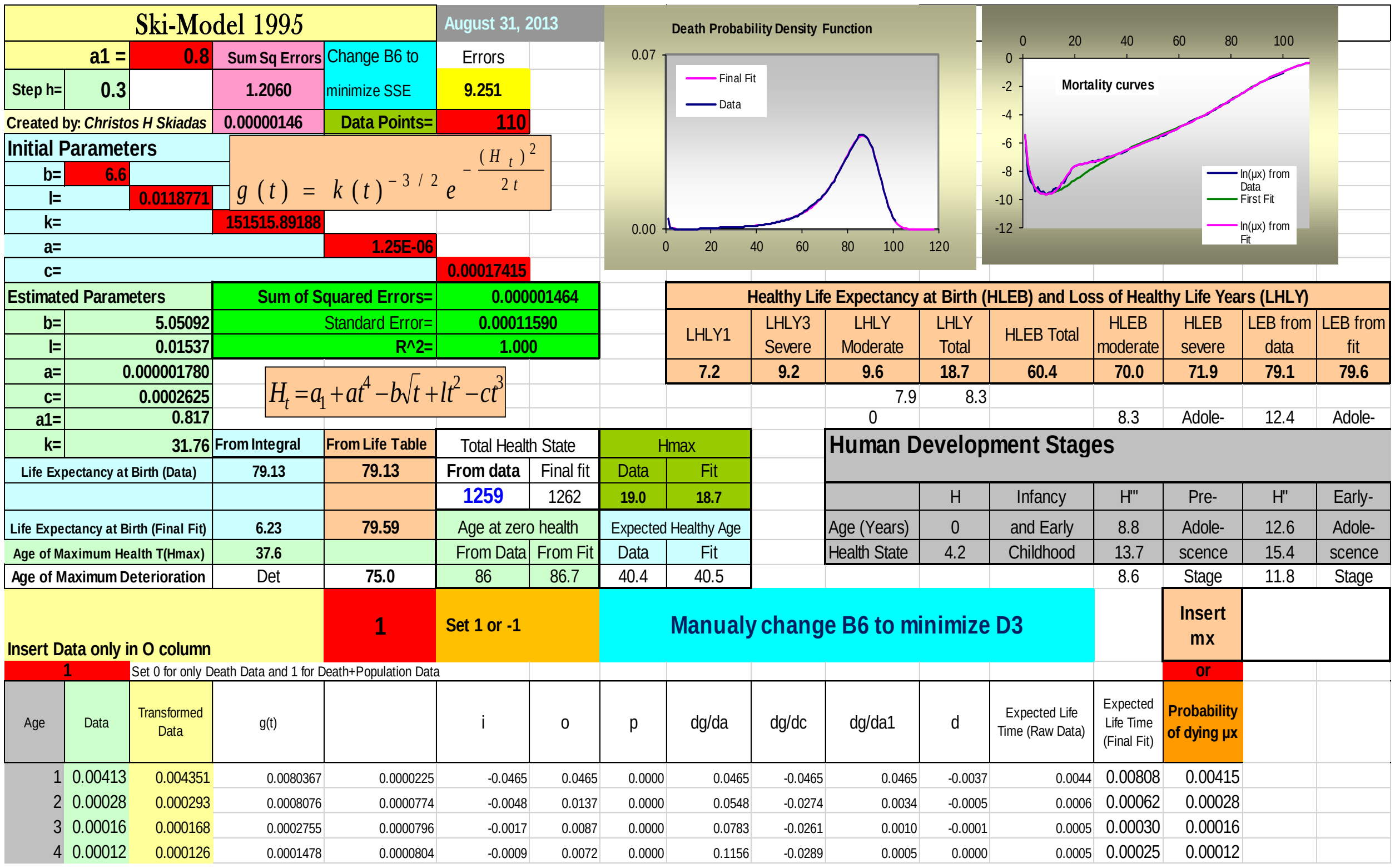

Fig. 10. Main view of the SKI-6 Excel Program and the Healthy Life Years Lost estimates. Download from http://www.smtda.net/demographics2020.html

## Conclusions and Further Study

We have provided an analytic explanation of the behavior of a parameter bx related to the healthy life years lost. We have also present an analytic formulation for the observations done along with the development of the appropriate extensions of the classical life tables so that to give a valuable tool for estimating the Healthy Life Years Lost. We have also presented on how the Weibull model properties expressing the fatigue of materials and especially the cumulative hazard of this model can express the additive process of disabilities and diseases to human population. In this paper we further analytically derive a more general model of survival-mortality in which we estimate a parameter related to the Healthy Life Years Lost (HLYL) and leading to the Weibull model and the corresponding shape parameter as a specific case. We have also demonstrated that the results found for the general HLYL parameter we have proposed provide results similar to those provided by the World Health Organization for the Healthy Life Expectancy (HALE) and the corresponding HLYL estimates. An analytic derivation of the mathematical formulas is presented along with an easy to apply Excel program. A further extension of the Excel program based on the Sullivan method provides estimates of the Healthy Life Expectancy at every year of the lifespan for five different types of estimates that are the Direct, WHO, Eurostat, Equal and Other. Estimates for several countries are presented. It is also introduced a methodology to bridge the gap between the World Health Organization (HALE) and Eurostat (HLE) healthy life expectancy estimates. The latest versions of this program appear in the Demographics2020 Workshop website at http://www.smtda.net/demographics2020.html .

## References

Jagger, C., Van Oyen, H. and Robine, J. M. (1999). Health Expectancy Calculation by the Sullivan Method: A Practical Guide.

Eurostat (2019). Healthy life years (from 2004 onwards) https://appsso.eurostat.ec.europa.eu/nui/submitViewTableAction.do

Román, R., Comas, M., Hoffmeister, L., Castells, X. (2007). Determining the lifetime density function using a continuous approach. *J Epidemiol Community Health*, 61:923–925. doi: 10.1136/jech.2006.052639.

Skiadas, C.H. and Skiadas, C. (2014) The First Exit Time Theory applied to Life Table Data: The Health State Function of a Population and other Characteristics. *Communications in Statistics-Theory and Methods*, 43, 1985-1600.

Skiadas, C.H. and Skiadas, C. (2015). Exploring the State of a Stochastic System via Stochastic Simulations: An Interesting Inversion Problem and the Health State Function. *Methodology and Computing in Applied Probability*, 17, 973-982.

Skiadas, C.H. and Skiadas, C. (2018). *Exploring the Health State of a Population by Dynamic Modeling Methods*. The Springer Series on Demographic Methods and Population Analysis 45, Springer, Chum, Switzerland. https://doi.org/10.1007/978-3-319-65142-2 .

Skiadas, C.H. and Skiadas, C. (2018). The Health-Mortality Approach in Estimating the Healthy Life Years Lost Compared to the Global Burden of Disease Studies and Applications in World, USA and Japan in *Exploring the Health State of a Population by Dynamic Modeling Methods*. The Springer Series on Demographic Methods and Population Analysis 45, Springer, Chum, Switzerland, pp 67-124. https://doi.org/10.1007/978-3-319-65142-2_4 .

Skiadas, C.H. and Skiadas, C. (2018). The Health-Mortality Approach in Estimating the Healthy Life Years Lost Compared to the Global Burden of Disease Studies and Applications in World, USA and Japan. In *Exploring the Health State of a Population by Dynamic Modeling Methods*. Springer, Chum, Switzerland. https://doi.org/10.1007/978-3-319-65142-2_4 .

Skiadas, C.H. and Skiadas, C. (2018). *Demography and Health Issues: Population Aging, Mortality and Data Analysis*. The Springer Series on Demographic Methods and Population Analysis 46. Springer, Chum, Switzerland. https://doi.org/10.1007/978-3-319-76002-5 .

Skiadas, C.H. and Skiadas, C. (2019). Relation of the Weibull Shape Parameter with the Healthy Life Years Lost Estimates: Analytic Derivation and Estimation from an Extended Life Table. *ArXiv* https://arxiv.org/ftp/arxiv/papers/1904/1904.10124.pdf .

Skiadas, C.H. and Arezzo, M. F. (2018). Estimation of the Healthy Life Expectancy in Italy Through a Simple Model Based on Mortality Rate. In *Demography and Health Issues: Population Aging, Mortality and Data Analysis*. Springer, Chum, Switzerland. https://doi.org/10.1007/978-3-319-76002-5_4

Skiadas, C.H. and Skiadas, C. (March 2019). Modeling the Health Expenditure in Japan, 2011. A Healthy Life Years Lost Methodology. arXiv:1903.11565.

Skiadas, C.H. and Skiadas, C. (April 2019). Relation of the Weibull Shape Parameter with the Healthy Life Years Lost Estimates: Analytic Derivation and Estimation from an Extended Life Table. arXiv:1904.10124

Sullivan, D. F. (1971). "A single index of mortality and morbidity." Health Services Mental Health Administration Health Reports **86**: 347-354.

Weibull, W. (1951). A statistical distribution function of wide applicability. *Journal of Applied Mechanics* 18, 3, 293-297.

WHO Life Expectancy and Healthy Life Expectancy Data provided in Excel. http://apps.who.int/gho/athena/data/GHO/WHOSIS_000001,WHOSIS_000015,WHOSIS_000002,WHOSIS_000007?filter=COUNTRY:*&format=xml&profile=excel